\documentclass{JINST}

\title{Fast 4$\pi$ track reconstruction in nuclear emulsion detectors based on GPU technology}
\author{A. Ariga$^a$\thanks{Corresponding author.}, T. Ariga$^a$ \\
\llap{$^a$} Albert Einstein Center for Fundamental Physics, Laboratory for High Energy Physics, University of Bern, \\
Sidlerstrasse 5, 3012 Bern, Switzerland\\
Email: \email{akitaka.ariga@lhep.unibe.ch}}

\abstract{Fast 4$\pi$ solid angle particle track recognition has been a challenge in particle physics for a long time, especially in using nuclear emulsion detectors. The recent advances in computing technology opened the way for its realization. 
A fast 4$\pi$ solid angle particle track reconstruction based on GPU technology combined with a multithread programming is reported here with a detailed comparison of processing time by CPUs with respect to using GPUs. 
By employing 3 state-of-the-art GPUs with a multithread programming, a 60 times faster processing of 3D emulsion detector data has been achieved with an excellent tracking performance in comparison with a single-thread CPU processing, corresponding to processing of 15 cm$^2$ emulsion surface scanned per hour.
}


\begin{document}

\section{Introduction}

The reconstruction of ionizing particle trajectories (tracking) is a fundamental subject in experimental particle physics. Advanced implementations of 3D tracking in terms of data processing speed have been achieved with nuclear emulsion detectors, which feature outstanding 3D spatial resolution of the order of nanometers. However, so far their potentialities have not been fully exploited due to the significant computation resources needed for the track reconstruction.

\begin{figure}[htbp]
	\center
	\begin{tabular}{cc}
	\includegraphics[height=5cm]{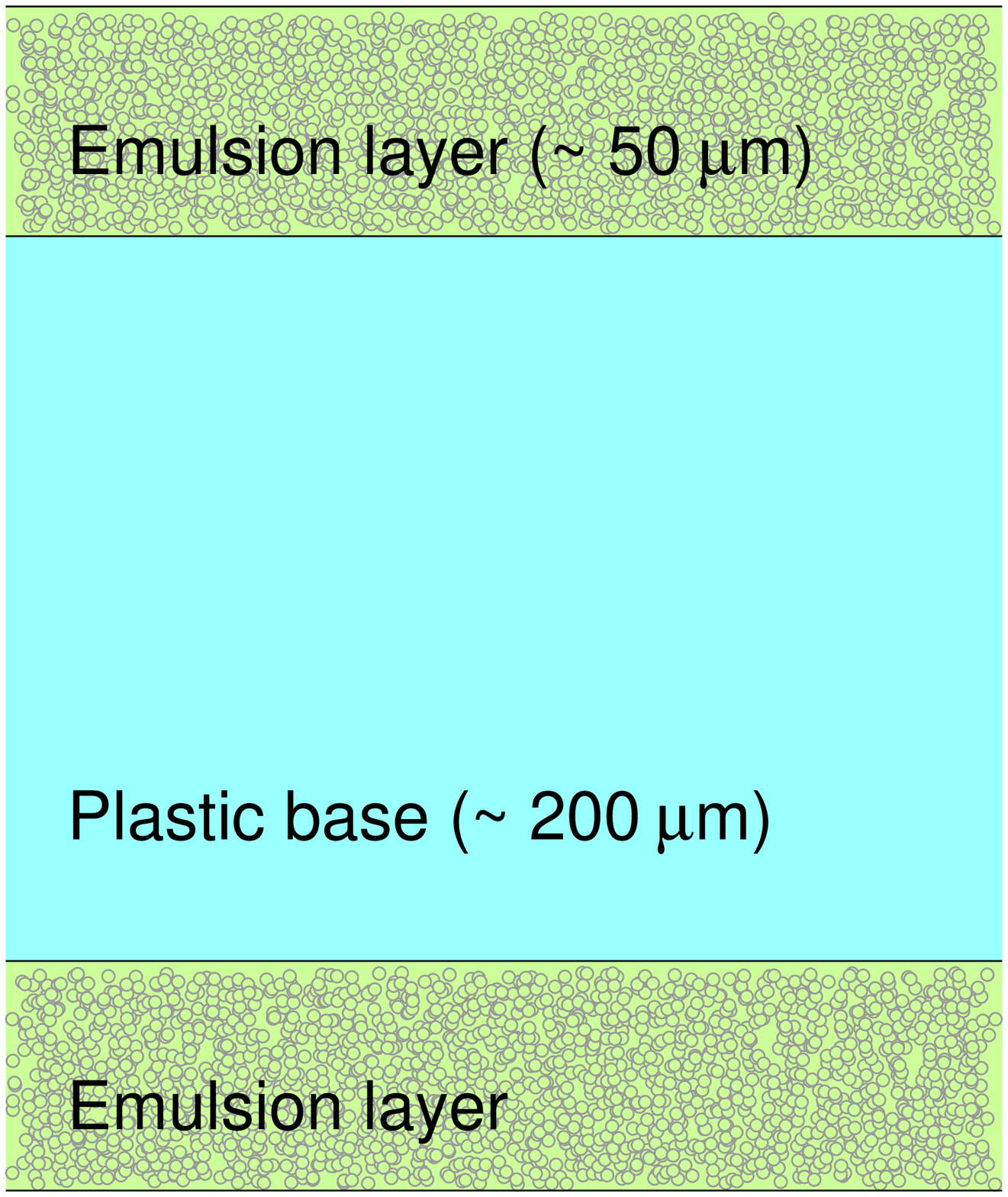}&
	\includegraphics[height=5cm]{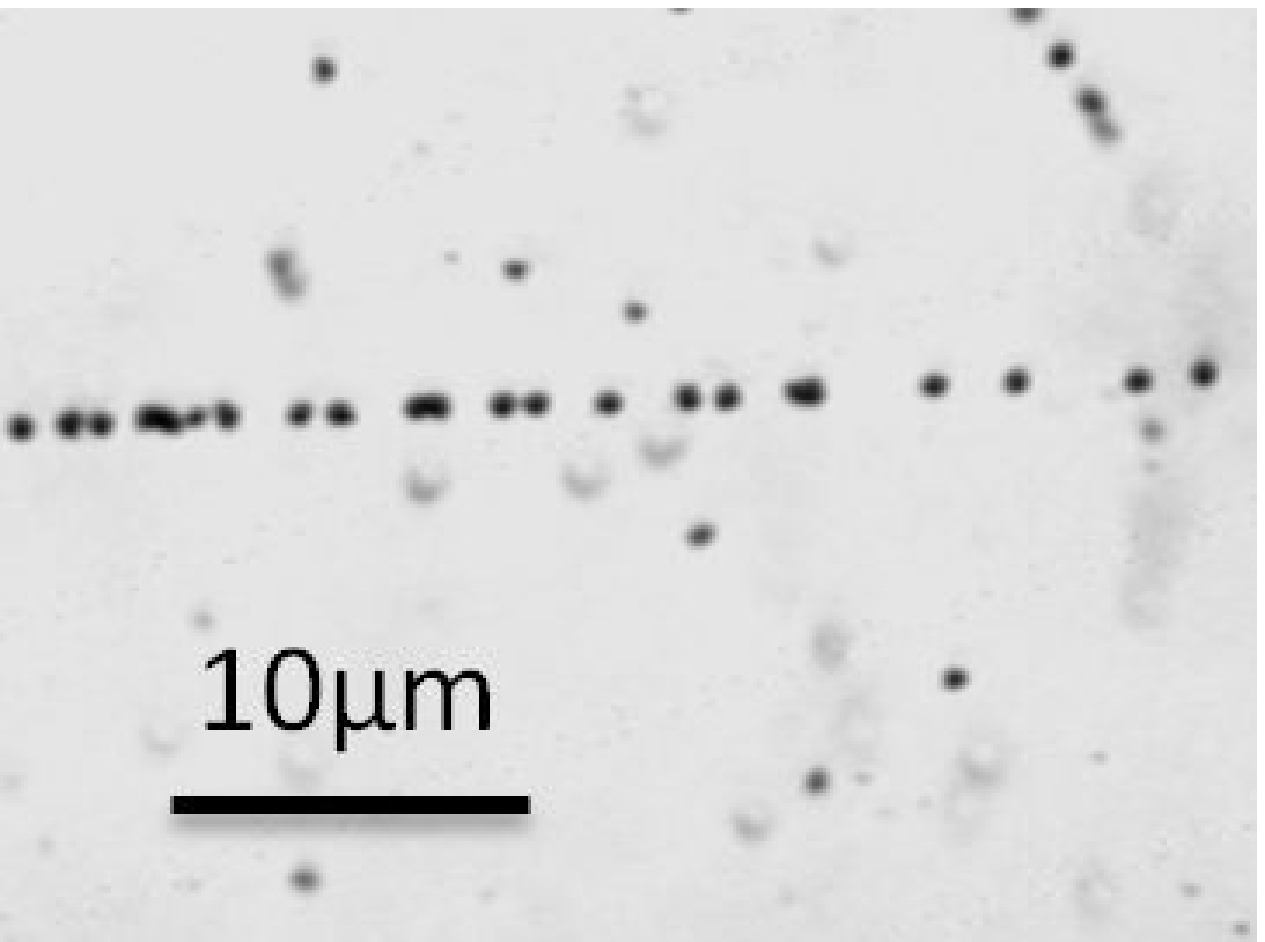}\\
	\end{tabular}
	\caption{Left: a schematic of typical emulsion detector structure. Right: a picture of a track after photo-development.}
	\label{fig:emulsion}
	\end{figure}

Nuclear emulsion detectors are made of silver-bromide microcrystals with a typical diameter of 200 nanometers uniformly distributed in a gelatin substrate \cite{emulsionreview, kuwabara}. Each microcrystal works as an independent 3D charged particle detector (micro-detector). A typical emulsion detector structure consists of two sensitive 50 $\mu$m thick emulsion layers deposited on both sides of plastic base $\sim$ 200 $\mu$m thick (see Figure \ref{fig:emulsion}-left) and has a surface of $\sim$ 10 cm $\times$ 10 cm, therefore such an emulsion detector features 1.3$\times$10$^{14}$ micro-detectors. Through a photo-developing process, the signals stored in the micro-detectors after the passage of an ionizing particle are chemically amplified and become visible under an optical microscope as dots, the so-called grains, with a diameter of 600 nanometers (see Figure \ref{fig:emulsion}-right). The readout and the tracking of signals from these micro-detectors have been performed so far by automated microscopes with a dedicated tracking system \cite{ts,s-uts,ess}. The raw data size from the detector is of the order of 10$^{13}$ Bytes (10 TByte) according to the number of micro-detectors. To reconstruct this amount of data in a reasonable time (some hours or less) a fast computing technology is therefore mandatory.


In the last decades, the speed of 3D tracking for nuclear emulsion detectors has remarkably improved with the evolution of electronics and computing technologies. The success of these implementations made possible the design and realization of the OPERA experiment \cite{opera}, which is the largest-ever project employing nuclear emulsions, aimed at the observation of flavor-changing neutrino oscillation in appearance mode. In order to deal with the large amount of data, in the OPERA experiment either a FPGA (Field-Programmable Gate Array) \cite{s-uts} or a  multi-CPU-thread computing \cite{ess} have been employed, achieving a surface scanning speed of as high as 20-50 cm$^2$/hour.

Nevertheless, due to the limited computing resources, the scanning systems of the OPERA experiment have large limitations as, for example, the poor angular acceptance in tracking or the difficulty in detecting short range tracks which start and stop in a single emulsion layer. Indeed, particles produced by high energy neutrino charge current interactions are boosted in the forward direction perpendicular to the emulsion surface (Z direction) and penetrate several emulsion layers. In this case the scanning system is optimized for track reconstruction at small angles ($\theta<0.5$ rad. $\theta$: track angle with Z axis)\footnote{The acceptance can be enlarged, however, since the computing time in the conventional algorithm has a linear dependence on the square of tangent of the angular acceptance $\theta_{acc}$, the angular acceptance is practically limited to $tan(\theta_{acc})=1$.} and for penetrating tracks, although nuclear emulsion detectors themselves can record particles at any direction. 

\begin{figure}[htbp]
	\center
	\begin{tabular}{cc}
	\includegraphics[height=4.5cm]{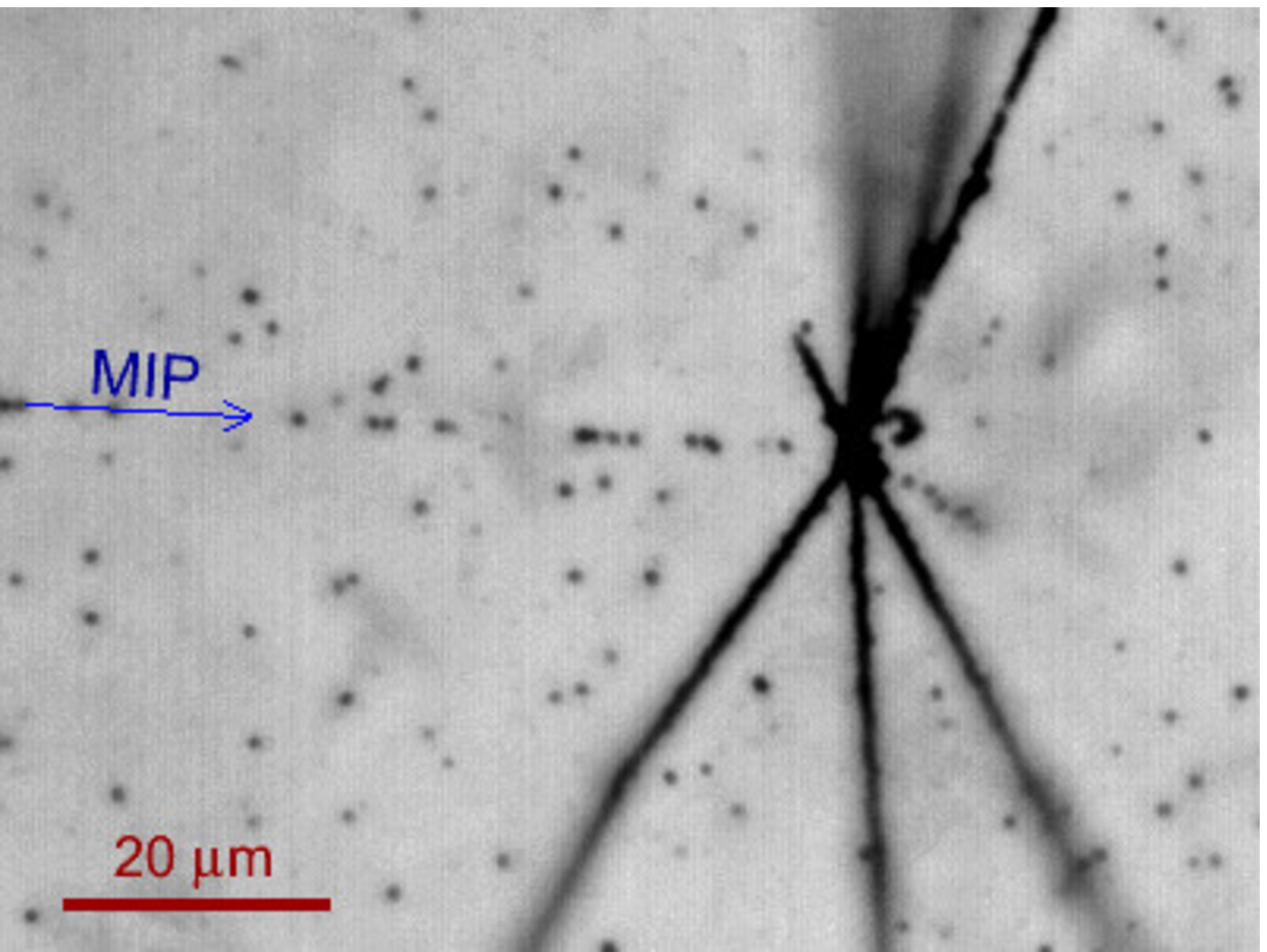}&
	\includegraphics[height=4.5cm]{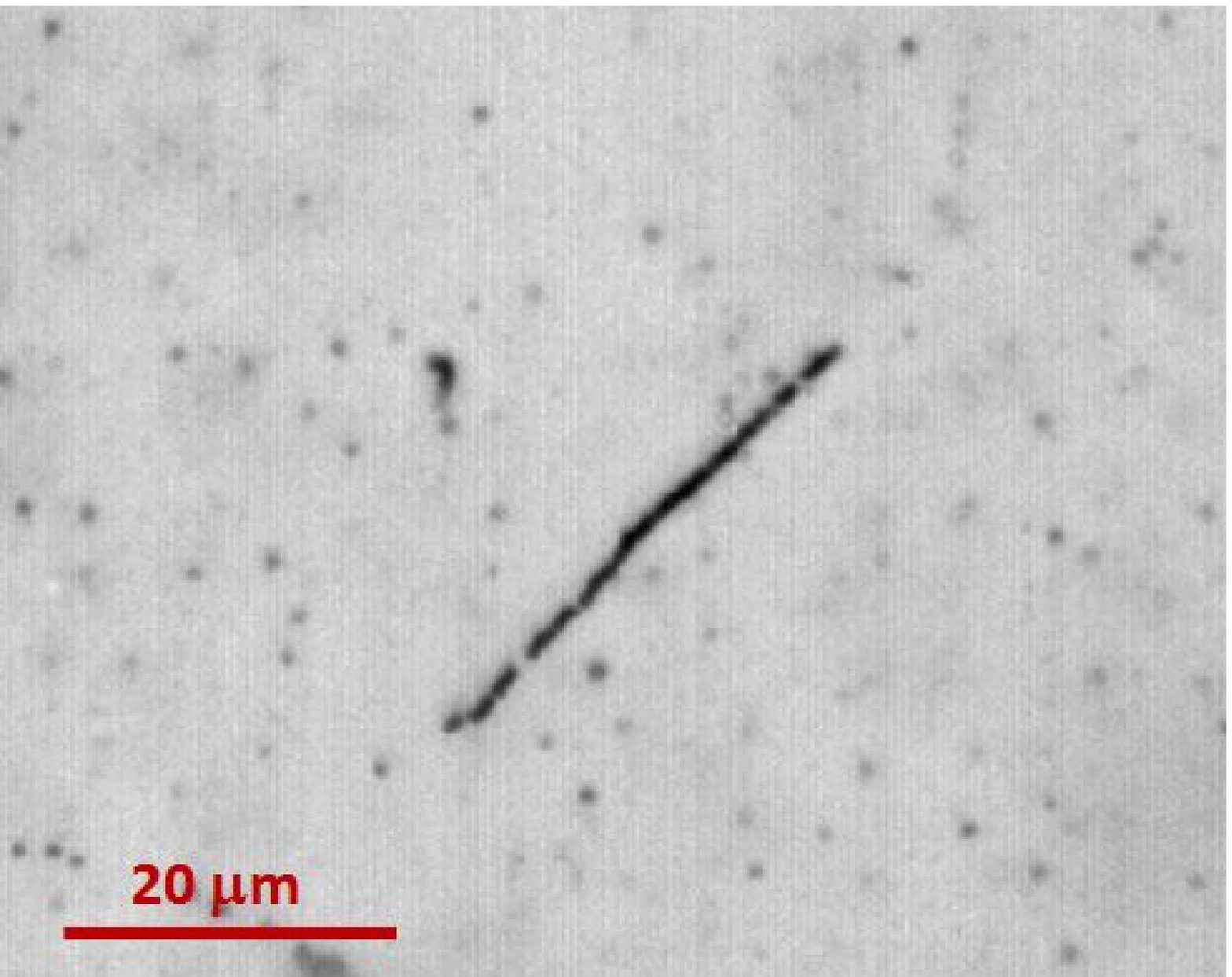}\\
	\end{tabular}
	\caption{Left: an antiproton annihilation vertex observed in the emulsion detector (from \cite{jinst2}). Right: a passage of proton recoiling from a 2.5 MeV neutron.}
	\label{fig:vertex}
	\end{figure}

However, such limitations make it difficult to use the same system in experiments with different particle topologies. As an example, Figure \ref{fig:vertex}-left shows an antiproton annihilation detected in one emulsion layer after a test performed for the future AEgIS experiment, which is aiming at the first measurement of the gravitational force on antimatter \cite{aegis}. Particles from the antiproton annihilation are basically emitted isotropically. In particular, the detection efficiency in the AEgIS conditions is estimated to be $\sim$ 60\% with no angular cuts on the particle directions, but only 2\% if the angular acceptance is the same as that of the OPERA scanning system \cite{jinst2}. Neutron flux monitoring is another example, where nuclear emulsion detectors are being used for dosimetry in medical applications or for nuclear fusion diagnostics. In these fields, the energy of neutrons is of the order of 1 MeV, which gives a range of recoiling proton of $\sim$ 10 $\mu$m (see Figure \ref{fig:vertex}-right). The OPERA tracking is blind to such short range tracks.
Therefore, in order to fulfill the needs of a potentially broader list of particle physics experiments not necessarily characterized by a forward particle topology as in OPERA, a 4$\pi$ directional fast track reconstruction turns out to be fundamental.

In the framework of this study, we propose a new tracking algorithm accomplishing the needs of a wide range of experiments. It requires about a factor 100 larger data processing than that of the OPERA experiment, while maintaining at the same time a high processing speed of the order of 10 cm$^2$/hour or more. 

We implemented the track reconstruction algorithm by using GPUs (Graphic Processing Unit). Each GPU has thousands of processing cores which can work in parallel. The use of GPUs for track reconstruction was already considered in 2003 by the Nagoya group. However, since the computing power was limited at that time and the programming infrastructure was not available for general purposes, the idea did not materialize into a real system. In 2006, NVIDIA introduced the CUDA infrastructure \cite{cuda} for general purpose computing with GPUs, which made possible the exploitation of the powerful computing power of GPU devices.

In this paper, we describe a new tracking algorithm for full solid angle track reconstruction in emulsion detectors by a GPU computing infrastructure.



\section{A new algorithm for particle track reconstruction}
\label{sec:algo}

In general, the aim of a tracking algorithm to be used for nuclear emulsion detectors is to recognize a consecutive sequence of 3D grains produced by the passage of charged particles.
A general purpose tracking algorithm should then perform as follows:
\begin{itemize}
\item reconstruction of particles tracks with no angular restriction;
\item identification of minimum and  heavily ionizing particles;
\item reconstruction of short-range particles starting and stopping inside a single emulsion layer.
\end{itemize}

\begin{figure}[htbp]
	\center
	\begin{tabular}{ccc}
	\includegraphics[width=4cm]{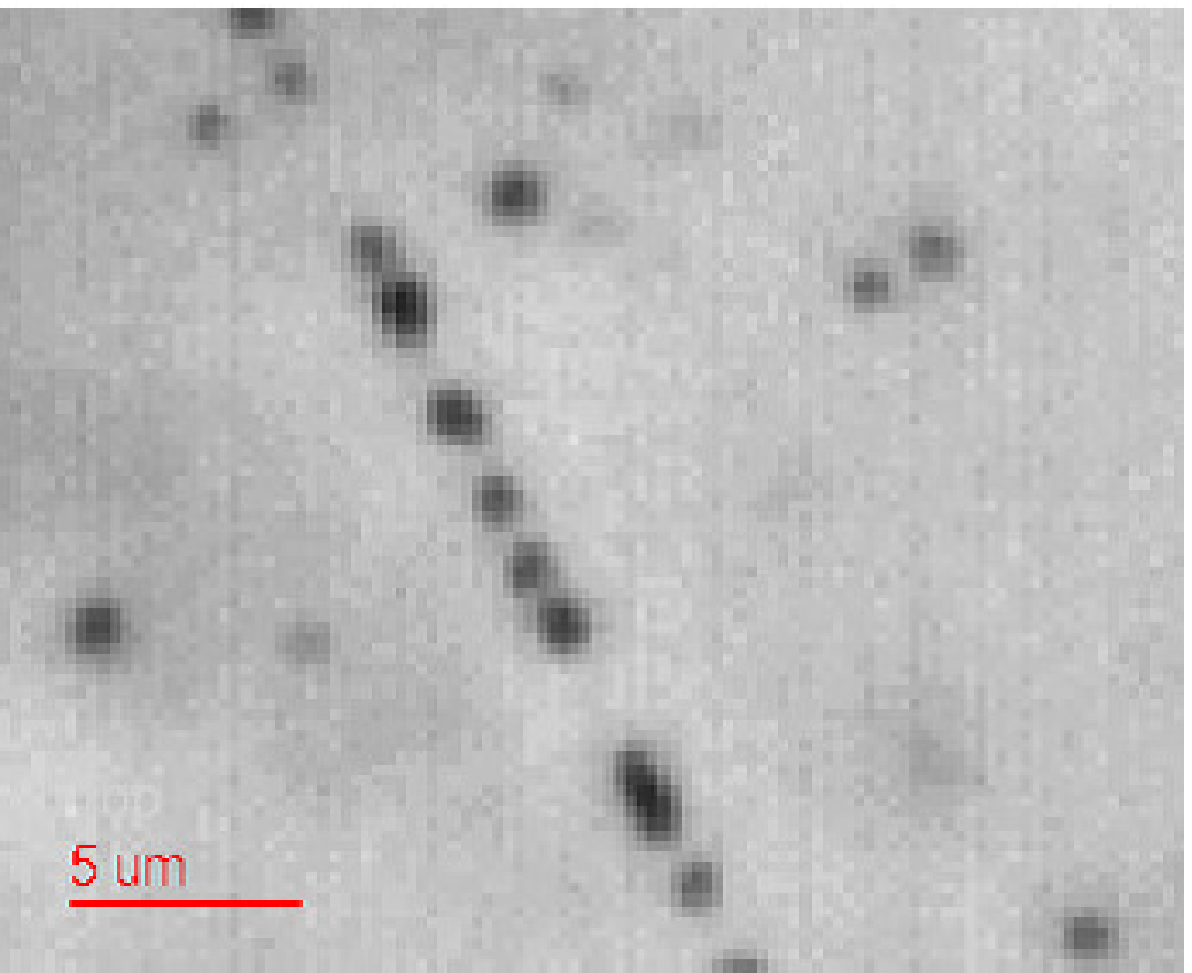}&
	\includegraphics[width=4cm]{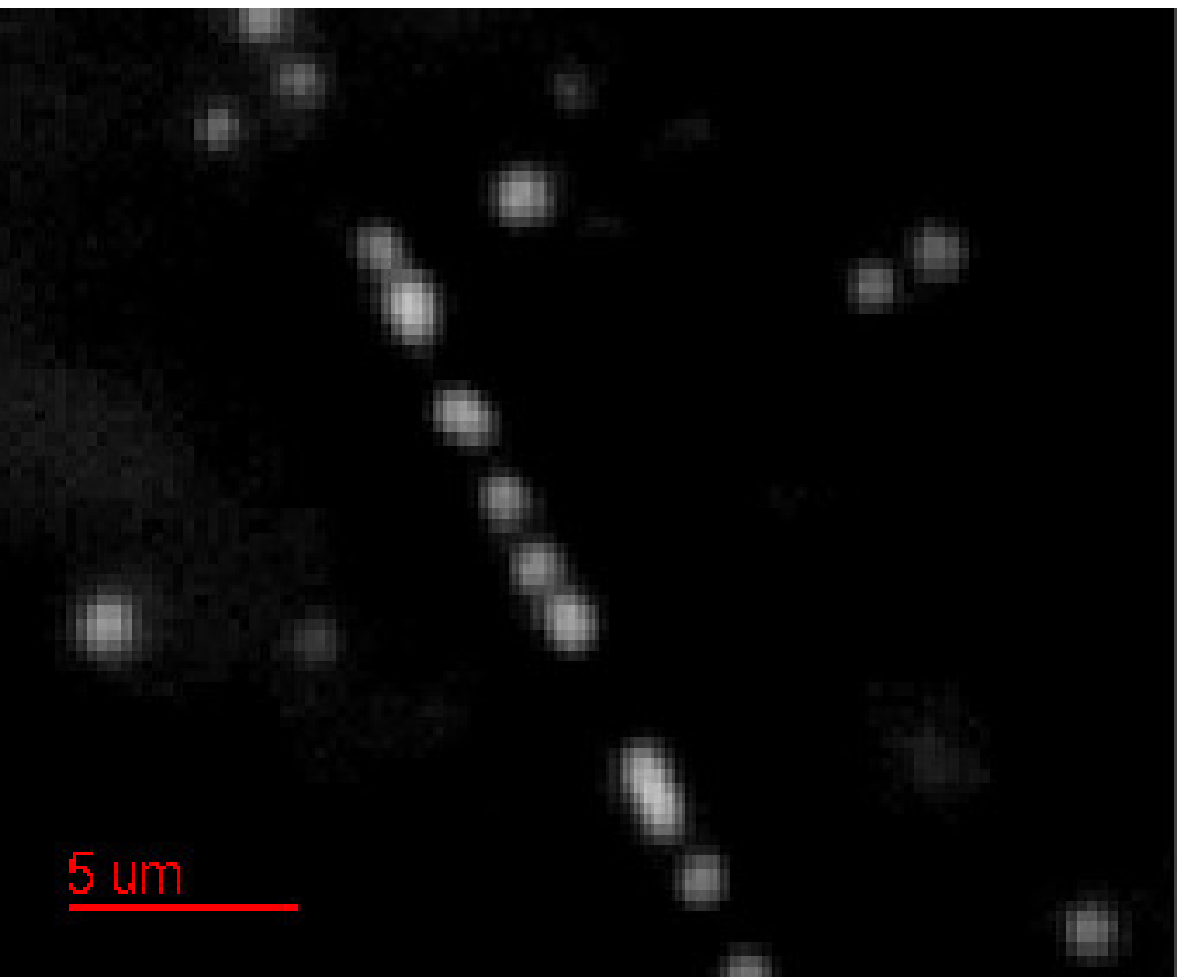}&
	\includegraphics[width=4cm]{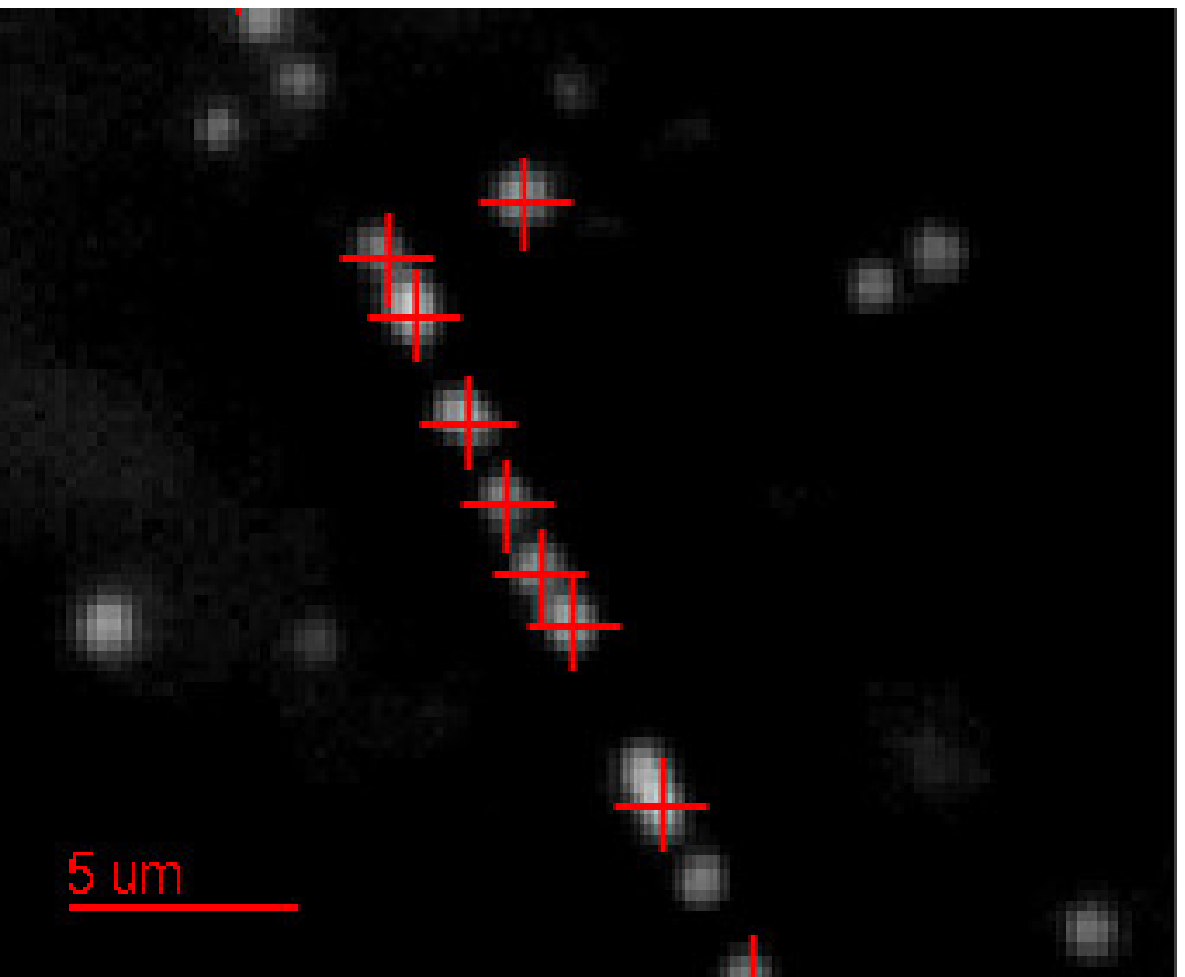}\\
	(a) & (b) & (c) \\
	\end{tabular}
	\caption{Processing of images. (a) Raw data image. (b) Filtered image. (c) Recognized grains marked with a red cross.}
	\label{fig:processing}
	\end{figure}

A basic unit of emulsion data called a $view$ corresponds to a 300 $\mu$m $\times$ 250 $\mu$m $\times$ 50 $\mu$m segment of the emulsion detector. Each view is digitized by a high speed CMOS camera into 1280 $\times$ 1024 pixels in X, Y and 40 frames in depth (Z). For the proposed tracking algorithm, a finer sampling of 40 frames, instead of the 16 frames used for conventional algorithms used for the OPERA experiment \cite{ess}, is chosen to allow short-range track reconstruction. A zoom of a typical frame is shown in Figure \ref{fig:processing}-(a). 

The pixels of the CMOS camera exhibit gain fluctuations, which cause the vertical lines seen in Figure \ref{fig:processing}-(a). In order to minimize this effect, a gain tuning and a threshold are applied on a pixel-by-pixel base. Also the image is inverted. These steps are called a $filtering$. An example of filtered image is shown in Figure \ref{fig:processing}-(b).

A grain produces a signal of a few pixels in X, Y and a few frames in depth. The 3D grain recognition is applied on the filtered image. Each pixel with a pulse height higher than a given threshold is compared to the $3\times 3\times 3$ neighboring pixels; the pixel with the highest pulse height among these will be defining the grain. The spatial coordinates of the grains are then computed with a center-of-mass method using the pulse height of each pixel (Figure \ref{fig:processing}-(c)). The 3D grain recognition is very critical especially for highly ionizing particles where the presence of a long optical shadow in the Z direction due to the high density of grains can, if the algorithm is not sufficiently intelligent, result in the incorrect recognition of grains. 

\begin{figure}[htbp]
	\center
	\begin{tabular}{cccc}
	\includegraphics[width=5cm]{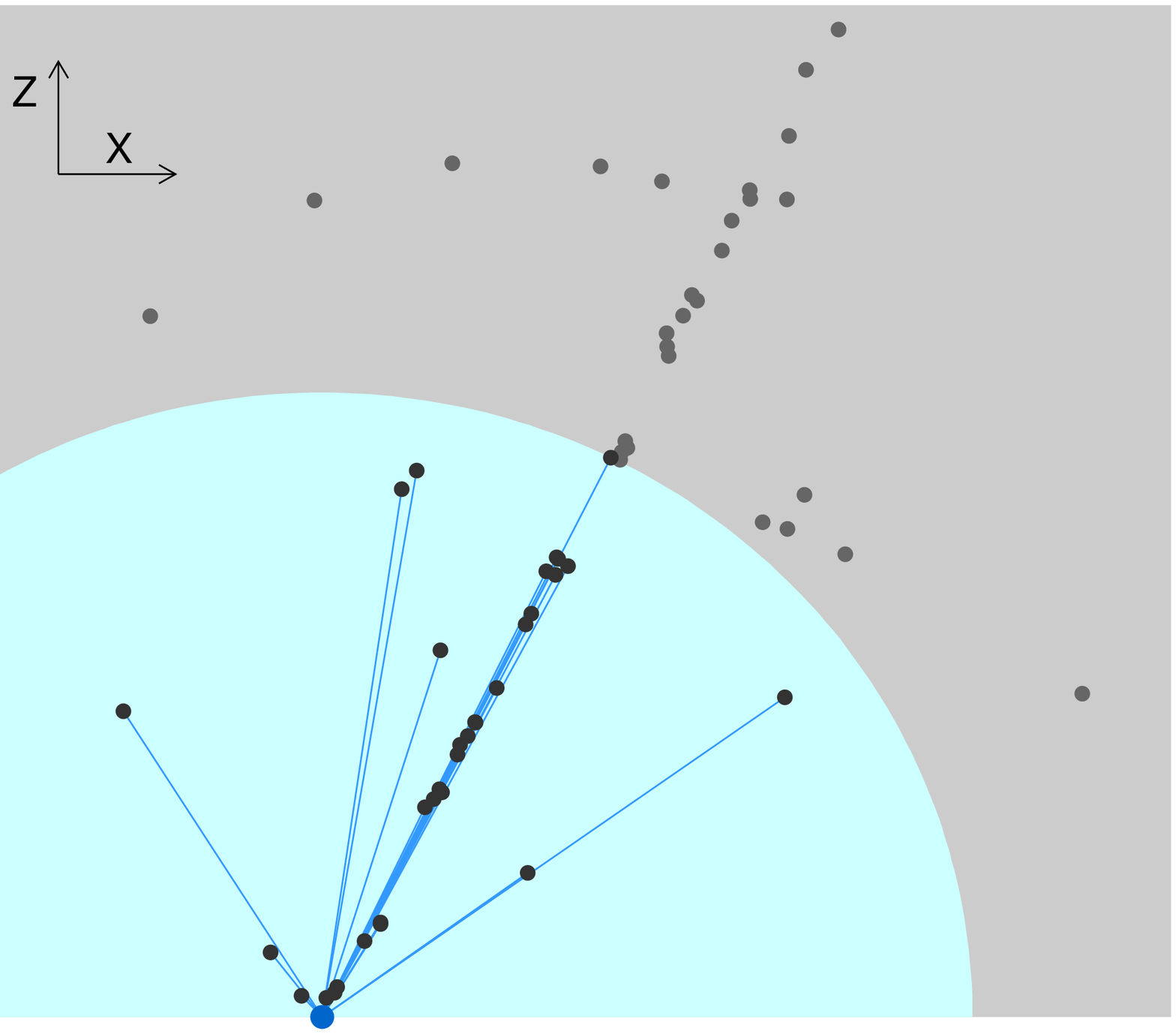}&
	\includegraphics[width=5cm]{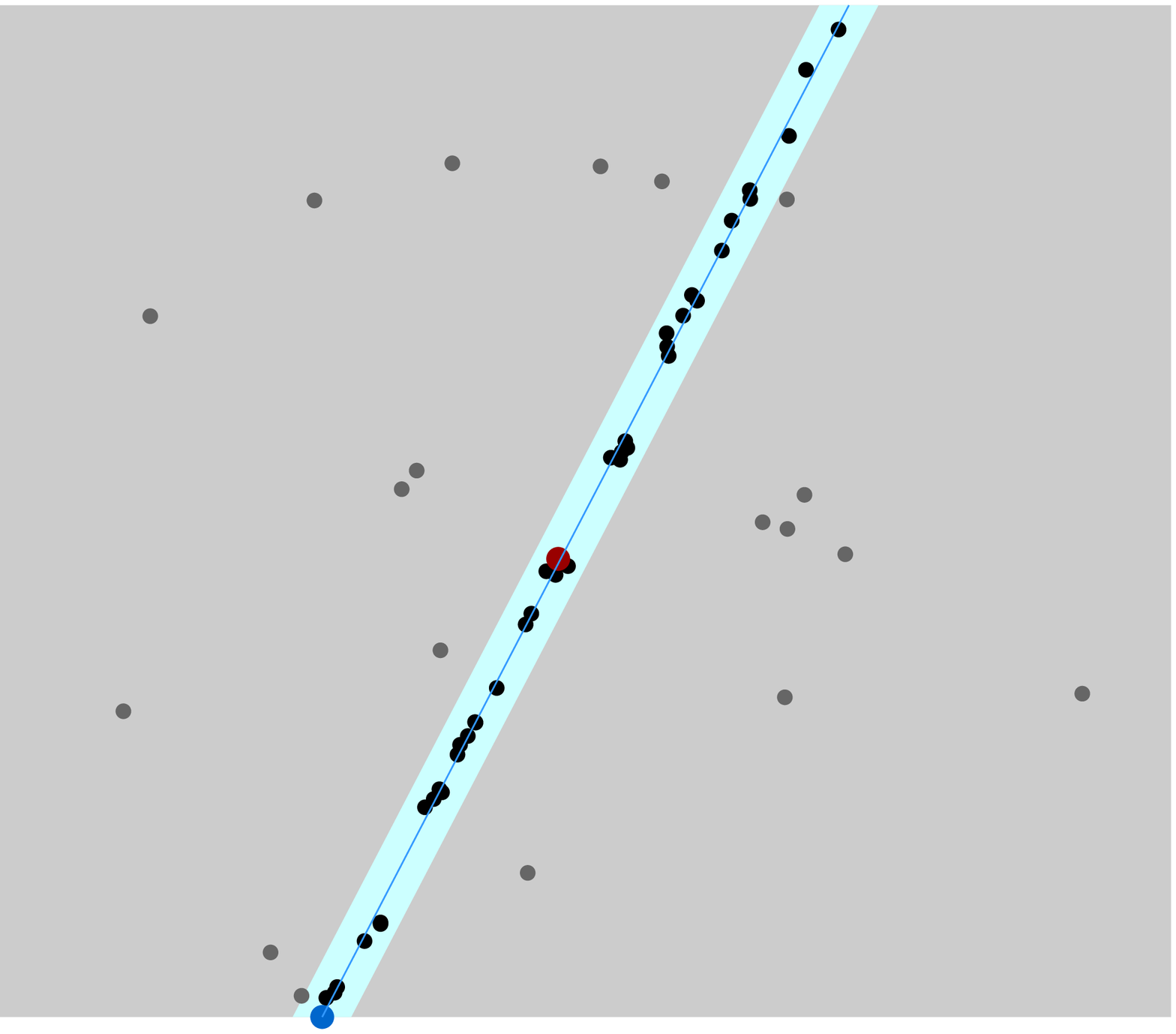}\\
	(a) & (b) \\
	\end{tabular}
	\caption{Tracking algorithm. (a) For each grain (for example, the blue dot), seeds for a track (blue lines) are defined by looping for grains within a given distance. (b) For each seed (blue line formed by blue and red dots), the number of grains along the seed is counted. If the number of grains is larger than a preset threshold, one can define it as a track.}
	\label{fig:tracking}
	\end{figure}

The seeds of tracks (lines) are defined as any grain pair whose distance is smaller than 25 $\mu$m, covering the 4 $\pi$ solid angle (Figure \ref{fig:tracking}-(a)). For each seed, a line is then defined connecting the two grains and the number of grains along the line is evaluated (Figure \ref{fig:tracking}-(b)). If the number of grains is above a preset threshold (e.g. 5), the seed is classified as a track. After selecting tracks each grain is exclusively associated with one track using that grain (each grain can be used only once) in order to suppress multiple tracks reconstructed for the same trajectory; and again the threshold on number of grains is applied to all tracks.

\begin{figure}[htbp]
	\center
	\begin{tabular}{ccccc}
	\includegraphics[height=5cm]{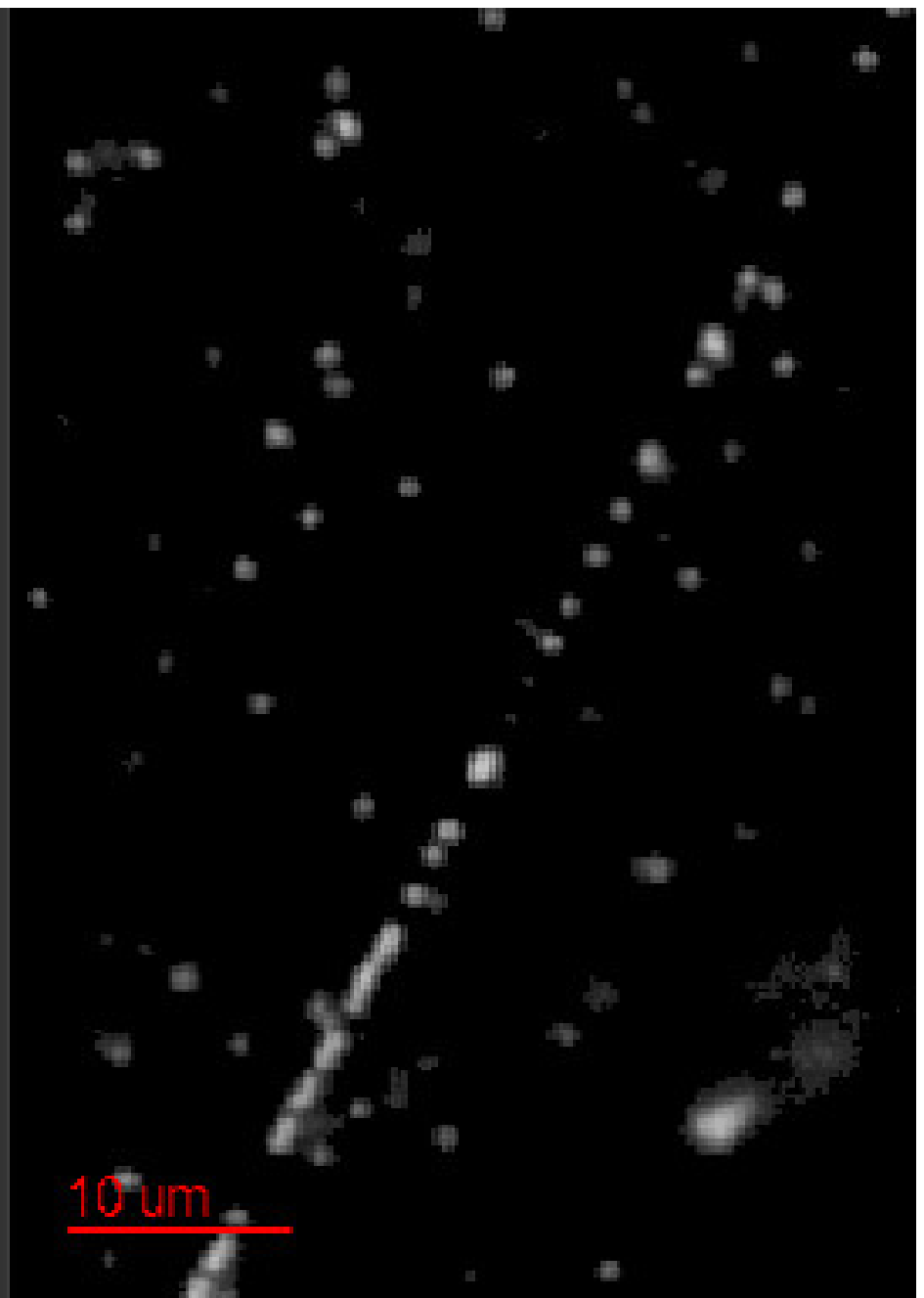}&
	\includegraphics[height=5cm]{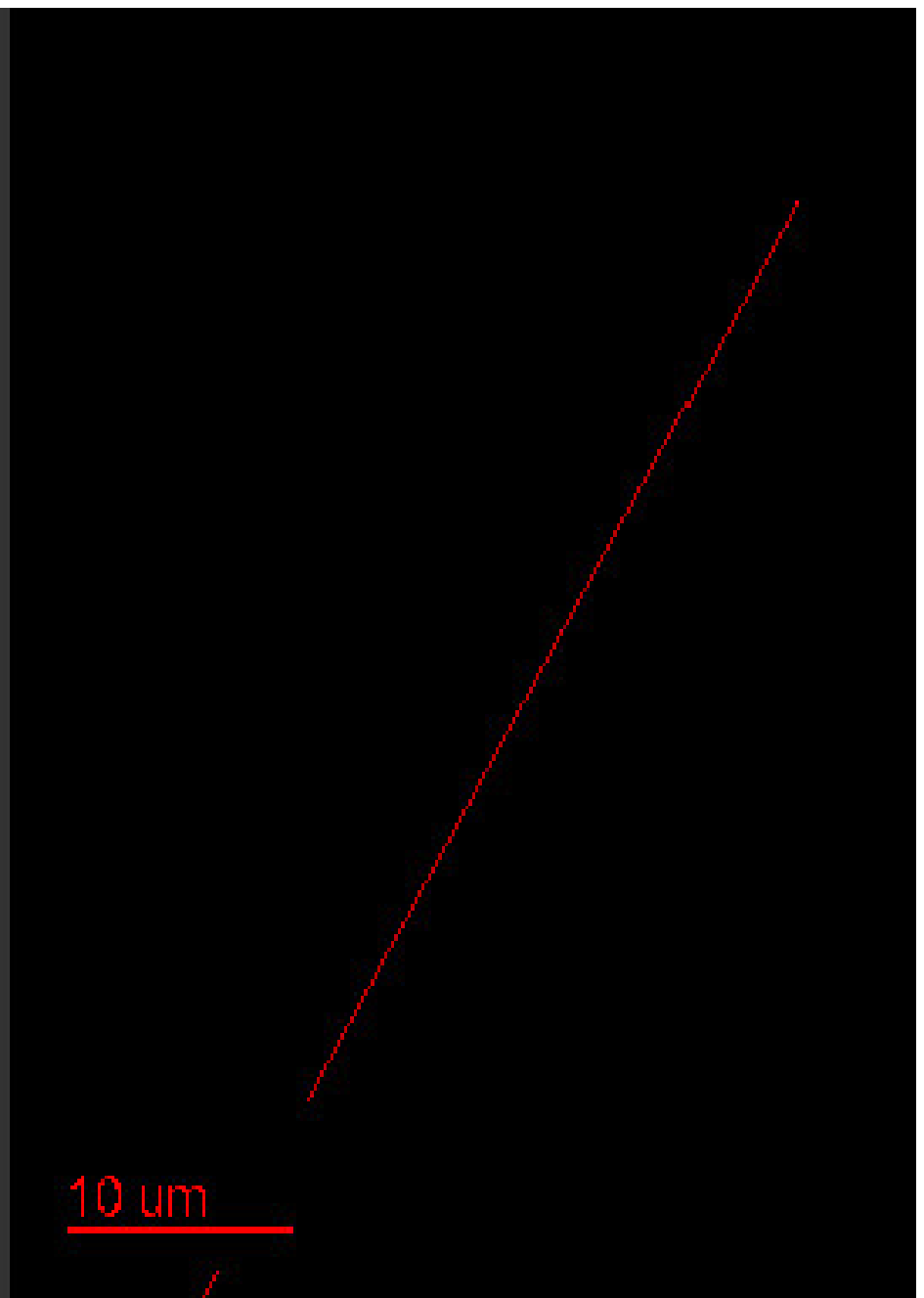}& &
	\includegraphics[height=5cm]{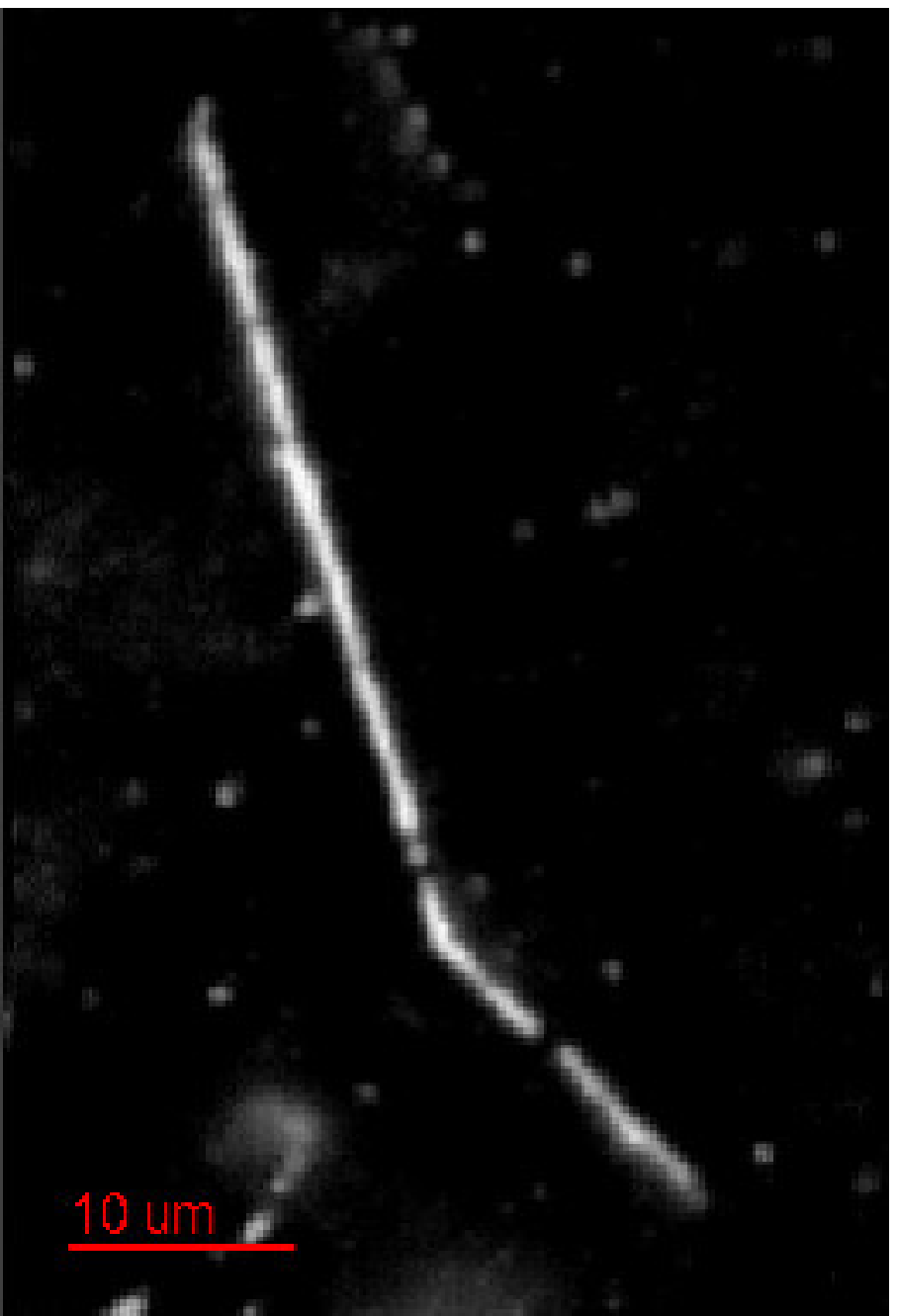}&
	\includegraphics[height=5cm]{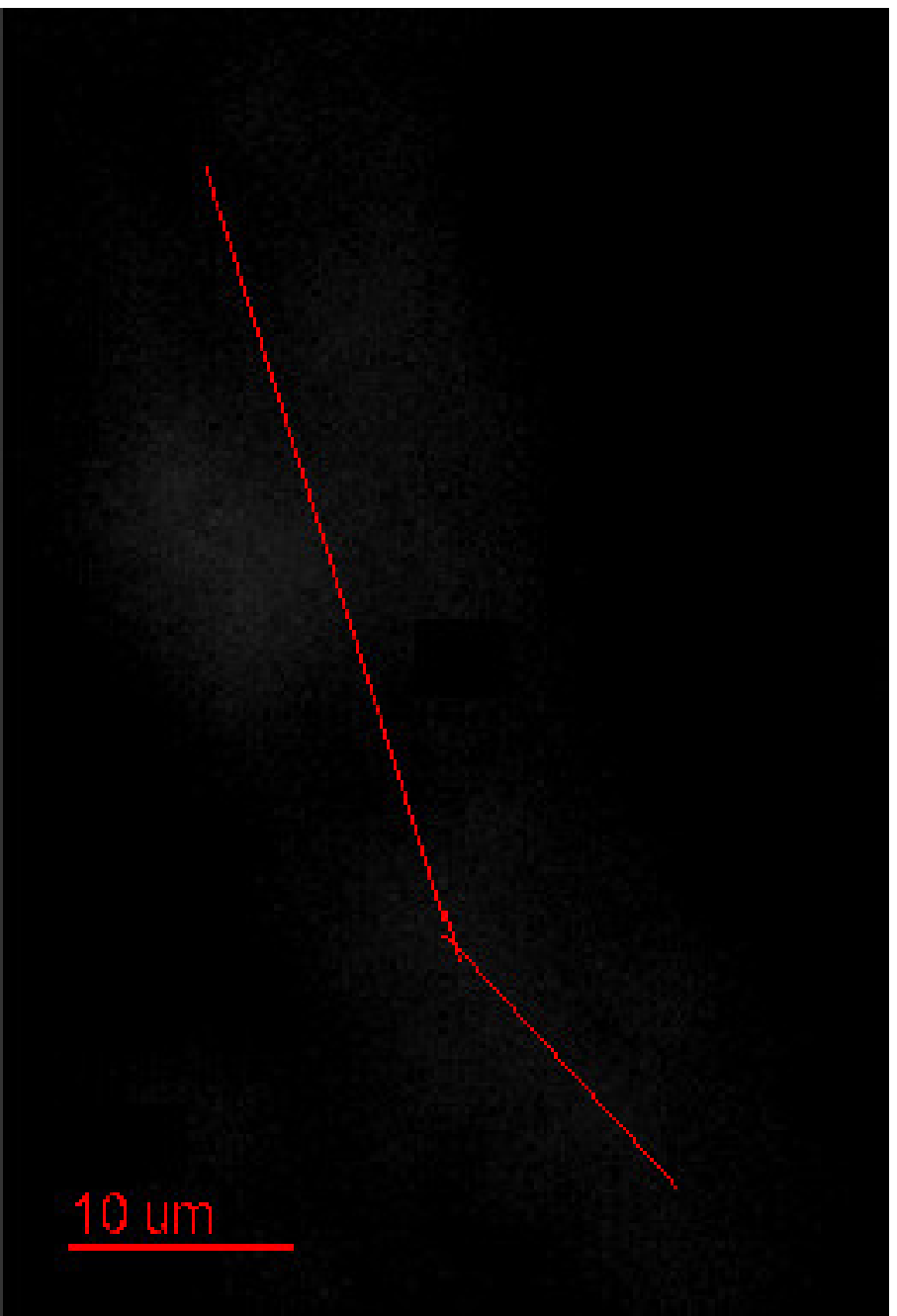}\\
	\multicolumn{2}{c}{(a) MIP track} & & \multicolumn{2}{c}{(b) Proton track recoiling from a 2.5 MeV neutron}\\
	\end{tabular}
	\caption{Example of tracks reconstructed by the proposed algorithm. (a) A MIP track running in parallel to the focal plane. (b) A proton track recoiling from a 2.5 MeV neutron and experiencing Rutherford scattering. For both cases, a filtered image is shown on the left and the reconstructed tracks (segments) are shown on the right as red lines, indicating the start and the end of the tracks. }
	\label{fig:tracks}
	\end{figure}

Examples of tracks reconstructed with this algorithm are shown in Figure \ref{fig:tracks}. The reconstruction of a cosmic-ray track is shown on the left, while a proton track recoiling from a 2.5 MeV neutron is shown on the right. Both the large angle MIP and the heavily ionizing particle are successfully reconstructed. A characteristic of the algorithm is the capability of reconstructing short tracks, and measuring whose length inside the emulsion. The trajectories of a proton before and after Rutherford scattering takes place are successfully reconstructed as two independent segments (Figure \ref{fig:tracks}-(b)-right).

The proposed tracking algorithm requires a larger amount of data processing with respect to the conventional codes. The angular acceptance of the latter is generally $\theta<0.5$ rad, as is the case for the OPERA experiment. The comparison of the angular acceptance for the new and the conventional algorithms is shown in Table \ref{tab:solidangle}. The coverage in solid angle is increased by a factor of 8. 
The detection of short tracks requires finer data taking and does not allow for fast tracking techniques used in conventional algorithm based on the assumption that particles enter from the top surface of the emulsion layer and escape from its bottom surface. The conventional code as described in \cite{ess} defines a limited number of pairs of frames that contribute to define seeds of tracks. With the proposed algorithm any combination of frames can be searched for.
The new condition increases the number of data processing by about 10 times.
From a simple multiplication of the above two factors (8$\times$10) it results that two orders of magnitude larger data processing capability are required to fully exploit the proposed algorithm.

\begin{table}
\center
\begin{tabular}{|l|r|r|}
\hline
   & New algorithm & OPERA \\
\hline
acceptance in $\theta$ (rad) & $\pi/2$  & 0.5 \\
\hline
acceptance in solid angle (steradian) & $4\pi$ & 1.5\\
\hline
acceptance efficiency & 100\% & 12\%\\
\hline
\end{tabular}
\caption{Comparison of angular acceptance between the new and the conventional tracking algorithms.}
\label{tab:solidangle}
\end{table}

\section{Track reconstruction based on GPU technology and performance}

In order to keep scanning speed high, we set the goal of a processing speed of 0.2 s/view on average, defined by the maximum data taking frequency of 5 Hz due to mechanical limitation of the microscopes and corresponding to a performance of 10 cm$^2$ of emulsion surface scanned per hour.

The implementation of the new algorithm described in Section \ref{sec:algo} using standard programming (single thread, C++ code) takes about 10 s/view with a state-of-the-art PC. The code already includes some basic fast computing procedures, such as the use of three dimensional hash tables\footnote{A view consists of 40 frames, and each frame is divided in cells of $32\times 32$ pixels in X and Y. Thus a view is divided into 51200 3D cells (or hash tables). All recognized grains are then attributed to these cells. In this way, the processing time is reduced as the following processes refer only to the grains in the relevant cells and not to all the grains of the view.}.

\begin{figure}[htbp]
	\center
	\includegraphics[width=11cm]{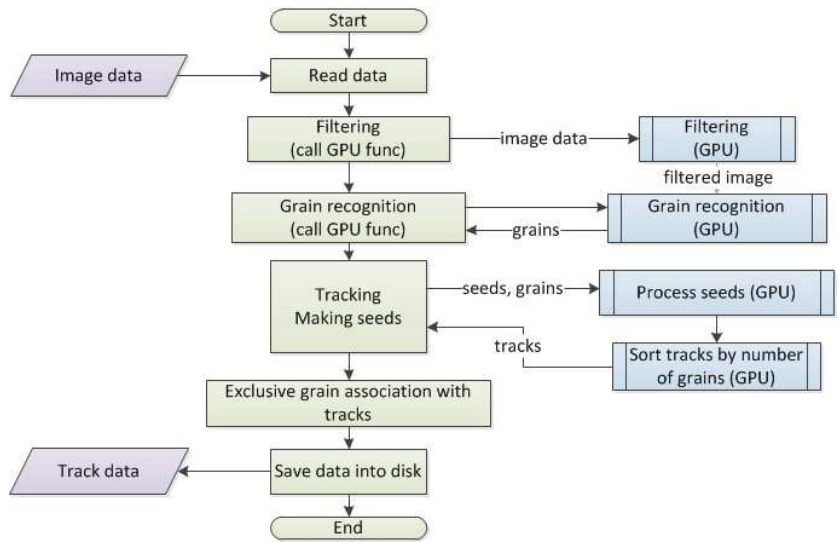}
	\caption{Flowchart of tracking procedure. Functions on the right (blue) are executed on GPUs.}
	\label{fig:flowchart}
	\end{figure}

The GPU programming has the advantage of parallel processing and it is very suitable for image processing and track reconstruction. The code is written with NVIDIA CUDA libraries (version 5.5) combined with CERN ROOT libraries \cite{root}. The algorithm is shared by CPU and GPU codes as shown in Figure \ref{fig:flowchart}. For an example, the process to form the seeds of tracks (described in Figure \ref{fig:tracking}-(a)) is performed by the CPU, while counting of grains along the seeds (Figure \ref{fig:tracking}-(b)) is accomplished by the GPU. Hundreds of seeds can be processed in parallel by the GPU.

 A dedicated processing server (Linux Mint 15) equipped with a recent CPU (i7-3930K, 3.2 GHz, 6 cores, 12 logical processors \cite{i7}) with 3 state-of-the-art GPUs (NVIDIA GEFORCE TITAN, 837 MHz, 2688 cores, 6.144 GByte on-board memory \cite{titan}) and with a fast memory (DDR3-2400) has been tested for our purpose. A NVIDIA GEFORCE TITAN has a computing performance as high as 4.7 TFLOPS (Tera Floating-point Operation Per Second), which is large when compared to that of a i7-3930K (0.15 TFLOPS).

For the speed performance measurement, an emulsion layer was fully digitized and stored on a hard disk drive.
The results of a comparison of processing speed by a single CPU thread program with and without a GPU are shown in Table \ref{tab:singlethread}. The processing time per view is computed by averaging the processing time for 60 views. It also includes the time required for the data transfer between the host memory and the GPU memory. Each process is remarkably accelerated by about a factor 10. The total processing time, dominated by the 3D tracking process, is reduced to 1/16. 

\begin{table}
\center
\begin{tabular}{|l|r|r|r|}
\hline
Process & CPU & GPU & Gain\\
        & (s/view)& (s/view) & \\
\hline
Image filtering      & 0.55  & 0.022 & $\times$25 \\
\hline
3D grain recognition & 0.20 & 0.025 & $\times$8 \\
\hline
3D tracking          & 5.90 & 0.373 & $\times$16 \\
\hline
Total processing time & 6.65 & 0.420 & $\times$16 \\
\hline
\end{tabular}
\caption{Comparison of processing time between the CPU and the GPU programs with a single CPU thread.}
\label{tab:singlethread}
\end{table}

On the basis of the single CPU thread programming, a multithread programming was then implemented. Each thread is responsible for processing a view. For the case of multithread with GPUs, each CPU thread is linked to one of the three GPU devices. By running several threads one can exploit all the GPU devices.

\begin{figure}[htbp]
	\center
	\includegraphics[width=16cm]{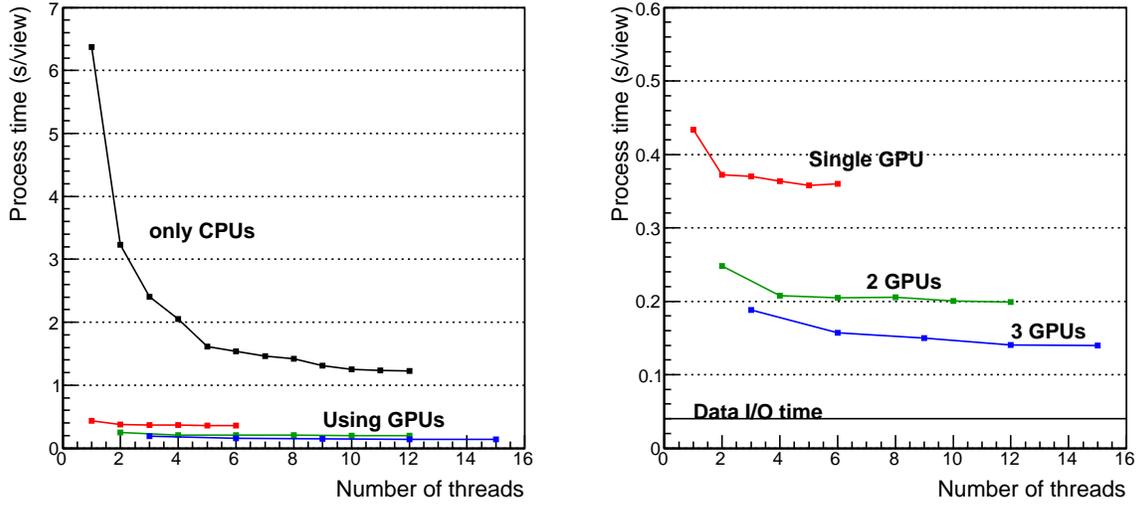}
	\caption{Comparison of processing time per view as a function of number of threads. Left: processing time by both the CPUs and the GPUs. Right: zoom for the GPUs.}
	\label{fig:multithread}
	\end{figure}

The comparison of the multithread programming based on the CPU and on the GPU is illustrated in Figure \ref{fig:multithread}. The processing time related to the CPU programming is 6.4 s/view with a single thread, and it is reduced as the inverse of the number of threads (black line in Figure \ref{fig:multithread}-left). However, for a number of threads larger than 6 no further gain is achieved in the processing time. This is due to the fact that the CPU has only 6 physical processors although it has 12 logical processors. The best result corresponds to 1.25 s/view with 12 threads. This achievement, however, is still far from our goal of 0.2 s/view.

On the other hand, with three GPUs and 12 threads (4 threads linked to one GPU), a processing speed of 0.14 s/view is achieved, out of which 0.04 s are needed for data I/O time. This corresponds to 15 cm$^2$ surface scanned per hour. The processing time is inversely proportional to the number of GPUs. However, processing with a number of threads per GPU larger than 2 is not beneficial in terms of speed. Eventually, the gain in the speed of processing time without data I/O from a single thread program based on the CPU with respect to the best multithread programming based on the GPU is about a factor of 60.
Currently the processing speed is limited by the memory access in the GPU and the memory transfer between GPU and host PC. Consequently there is still room for improvements.

\begin{figure}[htbp]
	\center
	\begin{tabular}{ccc}
	\hspace{-5mm}\includegraphics[width=5.5cm]{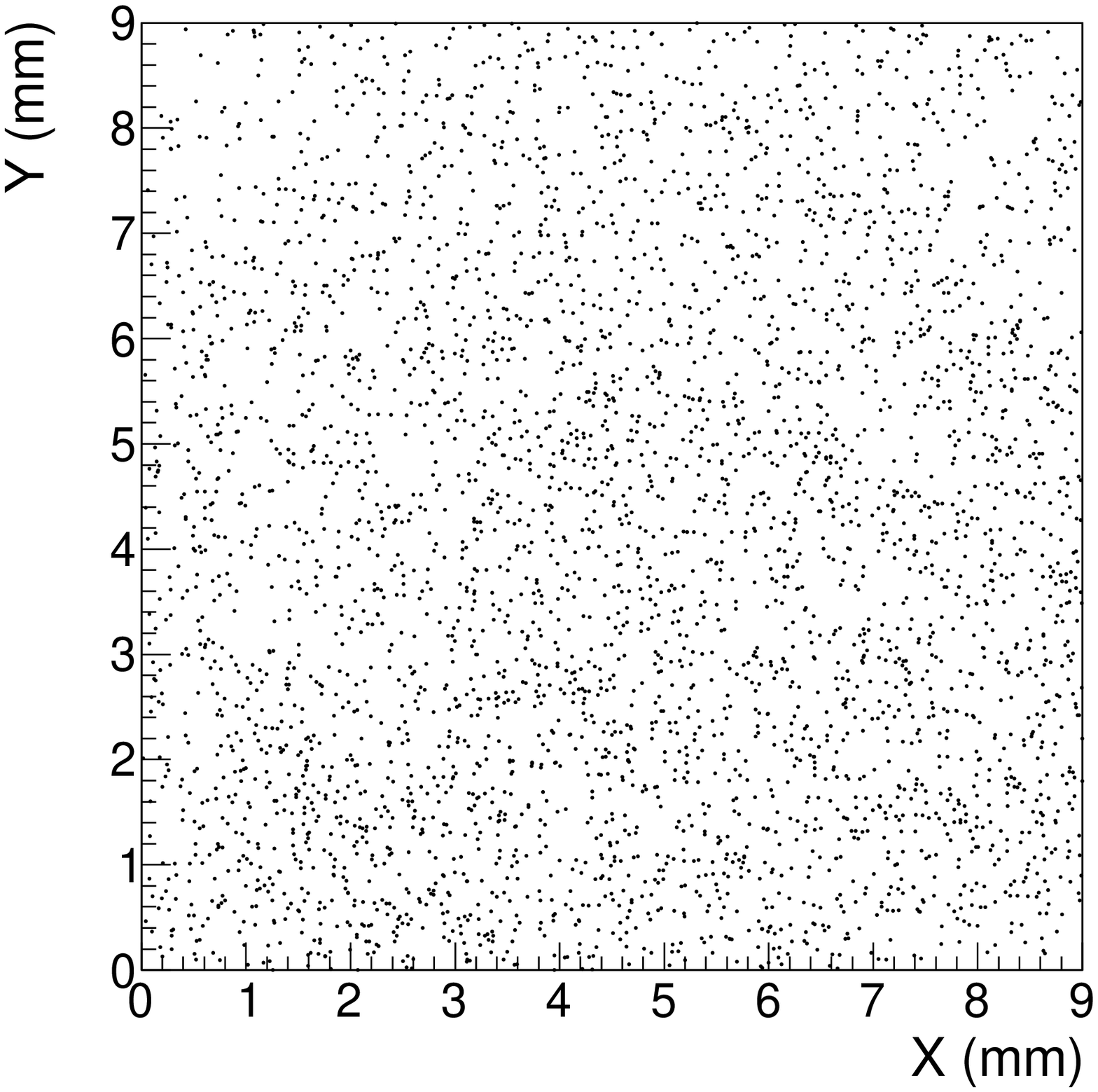}&
	\hspace{-5mm}\includegraphics[width=5.5cm]{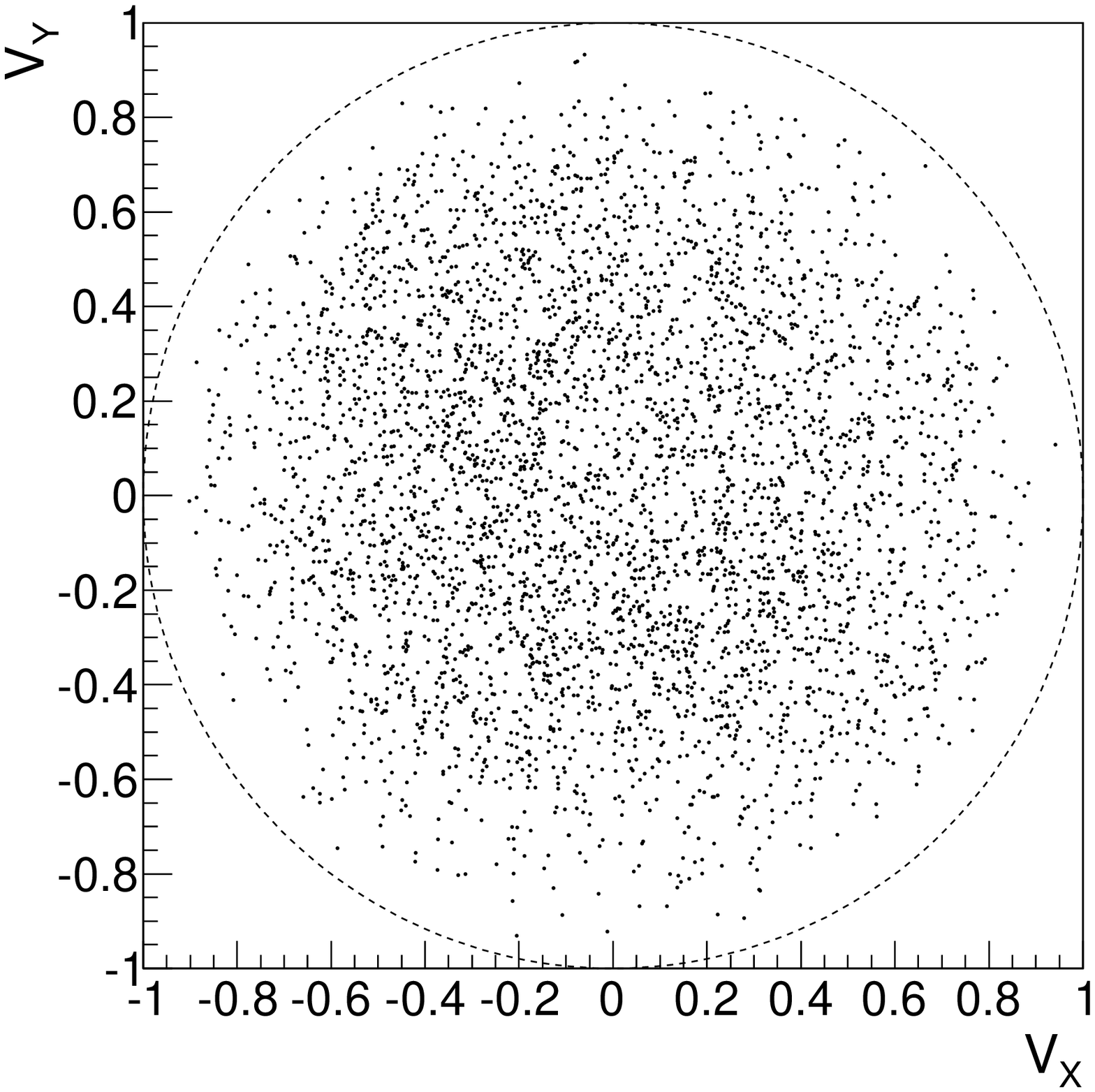}&
	\hspace{-5mm}\includegraphics[width=5.5cm]{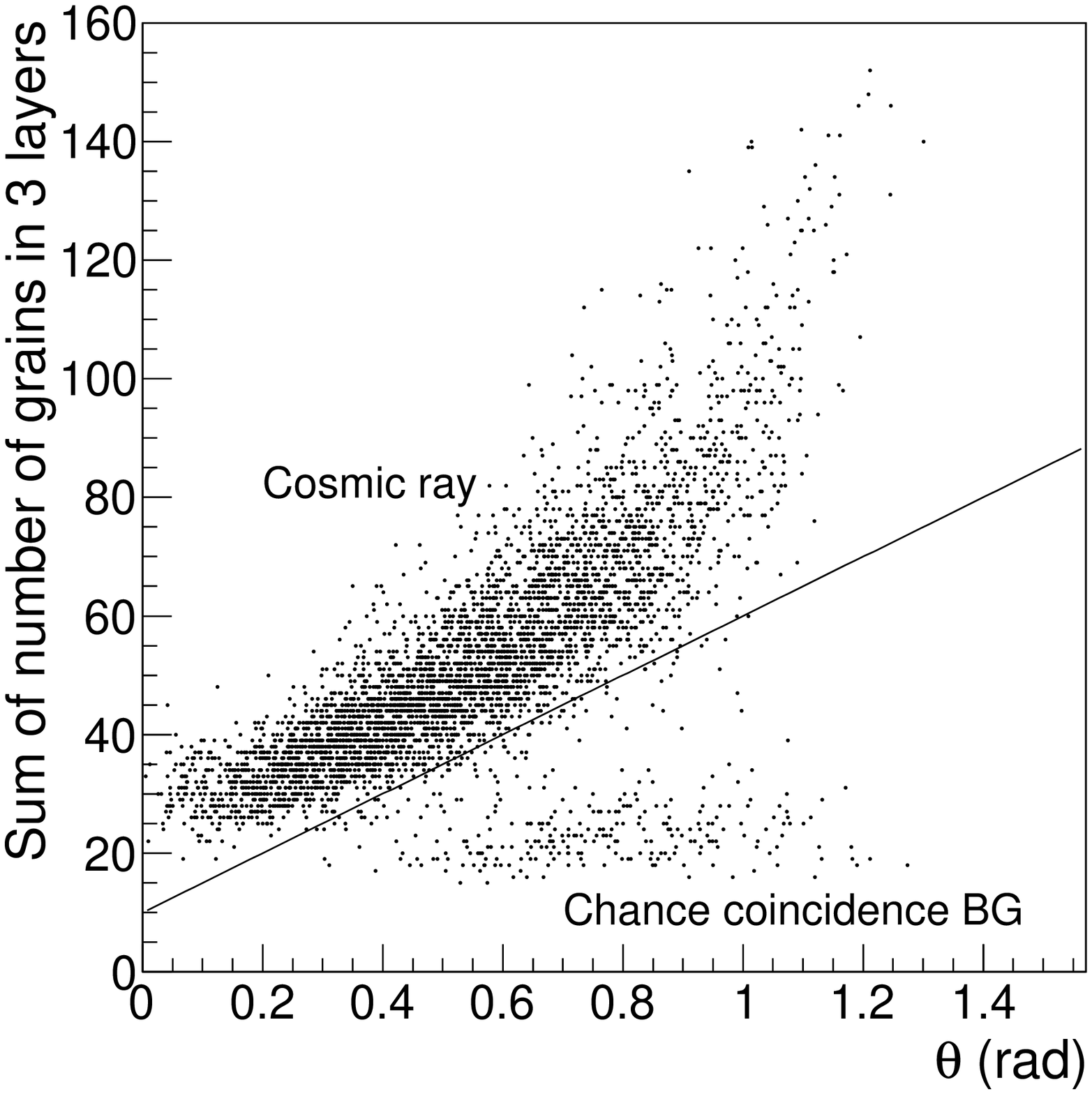}\\
	(a) & (b) & (c) \\
	\end{tabular}
	
	\caption{The distributions of the selected tracks for the efficiency study. (a) Track position distribution. (b) Track angular distribution. (c) Number of grains as a function of track angle.}
	\label{fig:cr}
	\end{figure}

\begin{figure}[htbp]
	\center
	\begin{tabular}{cc}
	\includegraphics[width=6cm]{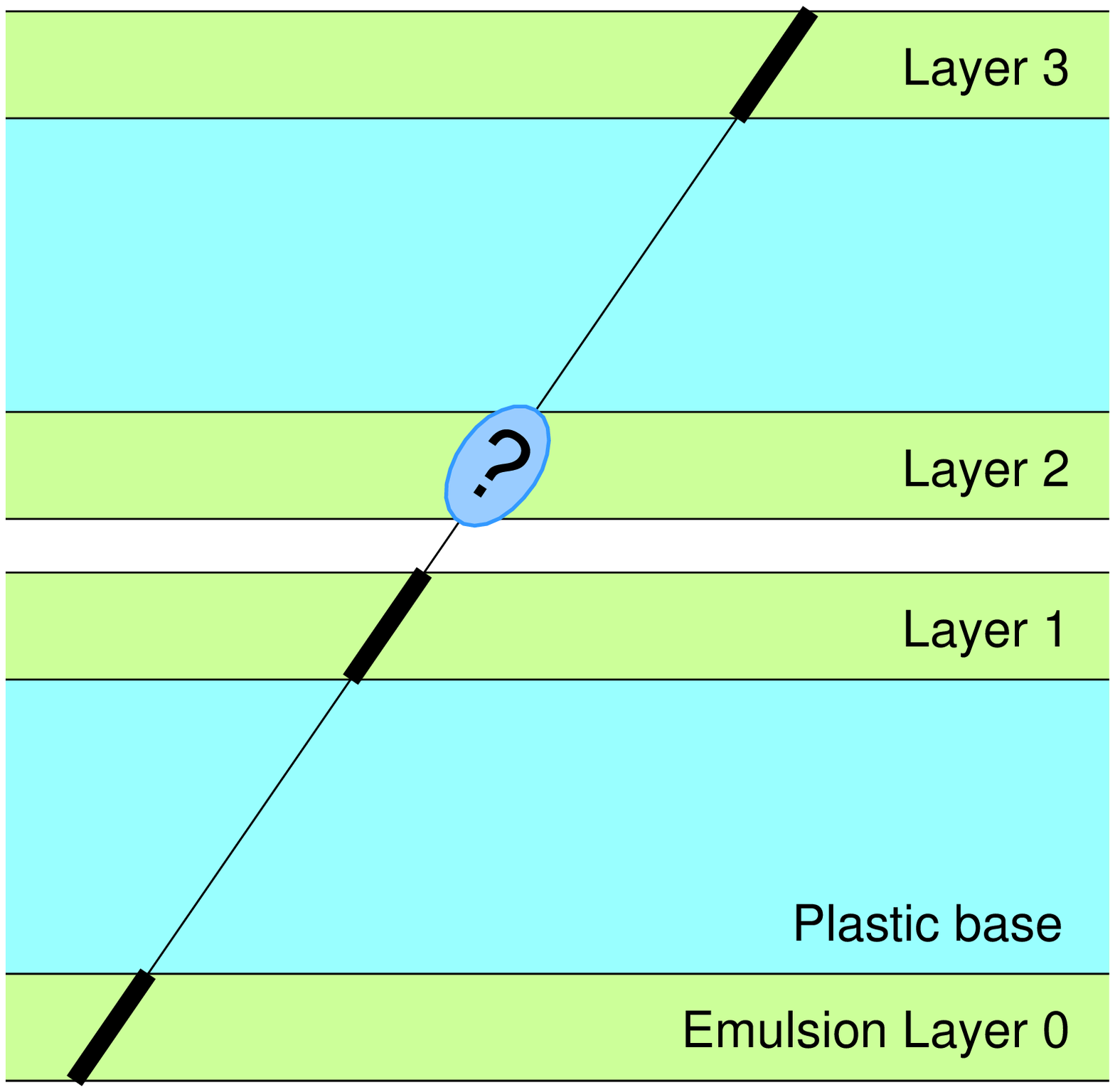}&
	\includegraphics[width=8.5cm]{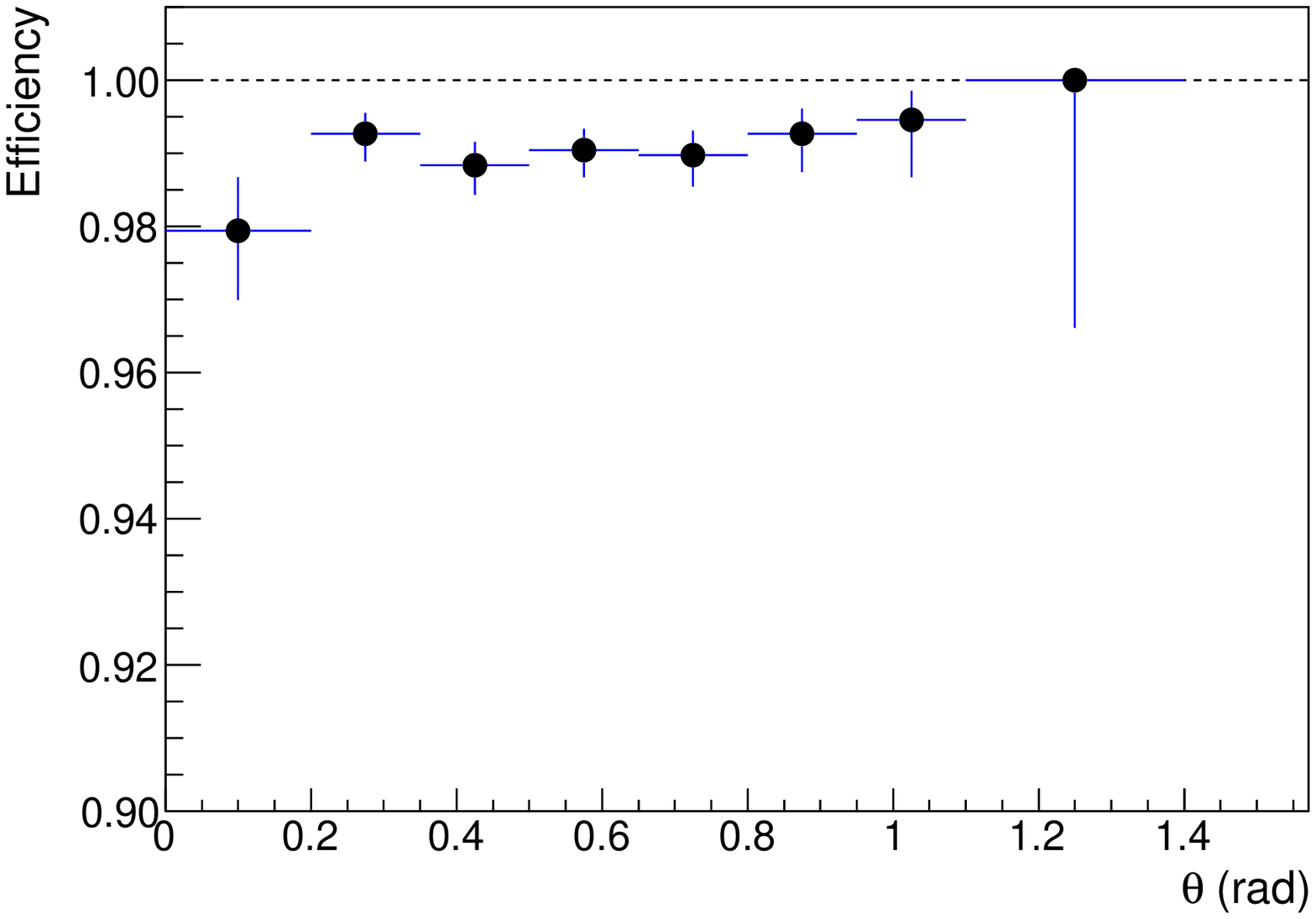}\\
	\end{tabular}
	\caption{A schematic of the tracking efficiency evaluation (left) and the tracking efficiency as a function of the track angle (right). The error bars on the tracking efficiency are obtained as Bayesian 68\% confidence bounds for binomial distributions, with a prior setting constant between 0 and 1, and to zero for the other region \cite{cowan}.}
	\label{fig:eff}
	\end{figure}

The efficiency of the track reconstruction algorithm was tested by using an emulsion detector module exposed to cosmic-rays (see Figure \ref{fig:eff} on the left). The emulsions are produced in the underground emulsion facility at the Laboratory for High Energy Physics (LHEP) of the University of Bern with an emulsion gel provided by the Nagoya University. The detector structure is made of two 50 $\mu$m-thick emulsion layers deposited on both sides of a 200 $\mu$m-thick plastic base. The sensitivity of the emulsion gel is measured to be 42$\pm$3 grains per 100 $\mu$m for MIP tracks. These are generated by 22 MeV electrons from the linac of the Clinic for Radiation Oncology of the Bern Inselspital. Two detectors are stacked together to form a module that was exposed to cosmic-rays for 1 week. An area of 1cm$^2$ was scanned for both detectors (a total of four sensitive layers) and the tracking algorithm was then applied to the microscope views. The alignment between the two detectors was performed by means of reconstructed cosmic tracks. As shown in Figure \ref{fig:eff}-left, the reconstructed tracks on the emulsion layers 0, 1 and 3 are selected to measure the tracking efficiency of the emulsion layer 2. Distributions of the selected tracks are shown in Figure \ref{fig:cr}. A flat track position distribution is evident from Figure \ref{fig:cr}-(a). The track angular distribution in terms of the X-Y components of a 3D unit vector is shown in Figure \ref{fig:cr}-(b). The tracks are reconstructed in a wide range of angles. The distribution of the number of grains is shown in Figure \ref{fig:cr}-(c). Larger angle tracks have consequently a higher number of grains as expected from geometrical considerations. Small contaminations from background due to random coincidences of low energy electron tracks are present in the bottom part of the distribution. In order to reject such a background, a cut on the number of grains is applied according to the line drawn in the figure.

The measured tracking efficiency is shown in Figure \ref{fig:eff}-right. The efficiency is found to be almost constant as a function of the track angle with a minor drop for small angles due to the reduced number of grains.
A very high average tracking efficiency of 99.0\% is measured.

The rate of fake reconstruction is also estimated by displacing data on the layer 2 by 1mm. The average fake rate is found to be 1.4\% and no strong angular dependence is observed. Thus the impurity in efficiency estimation, when a real track is missed and a fake track is found, results to be negligible ($\sim$0.01\% =(100\%-99.0\%)$\times$1.4\%).

\section{Conclusions}

Fast 4$\pi$ solid angle particle tracking has been for a long time a challenge when using nuclear emulsion detectors as high accuracy tracking devices. Advances in state-of-the-art computing technology have opened the way for the actual realization of fast 4$\pi$ tracking in emulsion detectors, as reported in this paper. A key element of our achievement has been a novel application of GPU technology.

Although the described tracking algorithm requires two orders of magnitude larger data processing as compared to conventional approaches, a satisfactory result assuring a processing performance of 15 cm$^2$ of emulsion surface scanned per hour is achieved by using the GPUs.
The gain in processing time by the multi-GPU with a multithread programming is measured to be more than a factor 60 with respect to the processing with a standard single thread programming. The tracking efficiency with the newly proposed algorithm is found to be as very high as 99\% for cosmic-ray tracks for a wide angular range.

This new tracking algorithm can be applied to any experiment making use of nuclear emulsion detectors which needs fast 4$\pi$ tracking, as the AEgIS experiment at CERN which needs high detection efficiency for antiproton annihilations. 
Furthermore the algorithm is being applied by us to several other applications, such as the measurement of proton tracks recoiling from neutrons and the cosmic-ray muon track radiography.

Similar approaches can also be applied to the reconstructions of other detectors like the liquid argon TPC and the silicon pixel detectors.

\acknowledgments
The authors wish to warmly acknowledge the colleagues at LHEP : C. Amsler, S. Braccini, A. Ereditato, J. Kawada, M. Kimura, I. Kreslo, L. Martinez, C. Pistillo, C. Regenfus, P. Scampoli, M. Schenk, J. Storey and S. Tufanli. We are also indebted to the colleagues of the Clinic for Radiation Oncology of the Bern Inselspital: M. Fix and P. Manser for making available the electron beam facility.


\begin{thebibliography}{9}

\bibitem{emulsionreview}
G. de Lellis, A. Ereditato and K. Niwa, Nuclear Emulsions, C.W. Fabjan and H. Schopper eds.,
Springer Materials, Landolt-B\"{o}ornstein Database (http://www.springermaterials.com),
Springer-Verlag, Heidelberg (2011).

\bibitem{kuwabara}
K. Kuwabara, S. Nishiyama, 
\emph{Development of New Nuclear Emulsion Film for Detection of Neutrinos by OPERA Experiment},
\emph{J. Soc. Photogr. Sci. Tech. Jpn.}, {\bf 67} (2004) 571.


\bibitem{ts} K. Niwa, K. Hoshino and K. Niu, 
\emph{Auto scanning and measuring system for the emulsion chamber}, in the proceedings of the International Cosmic ray Symposium of High Energy Phenomena, Tokyo, Japan (1974), see pag. 149.\\
S. Aoki et al., 
\emph{The fully automated emulsion analysis system}, 
\emph{Nucl. Instrum. Meth.} {\bf B 51} (1990) 466.\\
T. Nakano, 
\emph{Automatic analysis of nuclear emulsion},
Ph.D. Thesis, Nagoya University, Japan (1997).

\bibitem{s-uts} 
K. Morishima and T. Nakano,
\emph{Development of a new automatic nuclear emulsion scanning system, S-UTS, with continuous 3D tomographic image read-out},
\jinst{5}{2010}{P04011}

\bibitem{ess}
N. Armenise et al., 
\emph{High-speed particle tracking in nuclear emulsion by last-generation automatic
microscopes}, {\emph{Nucl. Instrum. Meth.} {\bf A 551} (2005) 261};

\bibitem{opera}  OPERA collaboration, M. Guler et al., 
\emph{An appearance experiment to search for $\nu_\mu \rightarrow \nu_\tau$ oscillations in the CNGS beam: experimental proposal}, \emph{CERN-SPSC-2000-028}, CERN, Geneva Switzerland (2000) [LNGS-P25-00].


\bibitem{aegis}
G. Drobychev et al., 
\emph{Proposal for the AEGIS experiment at the CERN antiproton decelerator},
available at spsc-2007-017.pdf;\\
AEGIS PROTO collaboration, \emph{Proposed antimatter gravity measurement with an antihydrogen beam},
{\emph{Nucl. Instrum. Meth.} \bf{B 266} (2008) 351}.


\bibitem{jinst2}
AEgIS collaboration, S. Aghion et al., 
\emph{Prospects for measuring the gravitational free-fall of antihydrogen with emulsion detectors},
\jinst{8}{2013}{P08013}.

\bibitem{cuda} NVIDIA CUDA web page:
\href{http://www.nvidia.com/cuda}{http://www.nvidia.com/cuda}.


\bibitem{root} CERN ROOT, \emph{A Data Analysis Framework}, 
web page:
\href{http://root.cern.ch/}{http://root.cern.ch/}.


\bibitem{i7} Intel$\textsuperscript{\textregistered}$  Core$^{\rm{TM}}$ i7-3930K Processor web page:
\href{http://ark.intel.com/products/63697}{http://ark.intel.com/products/63697}.

\bibitem{titan} NVIDIA GEFORCE TITAN web page:
\href{http://www.geforce.com/hardware/desktop-gpus/geforce-gtx-titan}{http://www.geforce.com/hardware/desktop-gpus/geforce-gtx-titan}.

\bibitem{cowan} 
G. Cowan, 
\emph{Statistical Data Analysis},
\emph{Clarendon Press, Oxford}, 1998.



%
%
%

\end{thebibliography}
\end{document}